# Nanostructured $LnBaCo_2O_{6-\delta}$ (Ln = Sm, Gd) with layered structure for Intermediate Temperature SOFC cathodes

Augusto E. Mejía Gómez [a,b], Diego G. Lamas [a,b,c], Ana Gabriela Leyva [b,c], Joaquín Sacanell [a,b*]

[a] CONICET, Argentina

[b] Departamento Física de la Materia Condensada, Centro Atómico Constituyentes, Comisión Nacional de Energía Atómica, San Martín, Pcia. de Buenos Aires, Argentina

[c] Universidad Nacional de Gral. San Martín, Escuela de Ciencia y Tecnología, San Martín, Pcia. de Buenos Aires, Argentina

**Abstract**

We evaluated for the first time the use of nanostructured layered perovskites of formulae $LnBaCo_2O_{6-d}$ with Ln = Sm and Gd (SBCO and GBCO, respetively) as SOFC cathodes, finding promising electrochemical properties in the intermediate temperature range (~700°C). The synthesis of these nanomaterials, not reported before, was achieved by using porous templates to confine the chemical reagents in regions of about 200 nm and 800 nm. The performance of nanostructured SBCO and GBCO cathodes for the oxygen reduction reaction was analyzed in symmetrical cells using $Gd_2O_3$-doped $CeO_2$ (GDC) as electrolyte. For this purpose, nanostructured SBCO and GBCO cathodes were deposited on both sides of the electrolyte by a simple thick-film procedure and evaluated by Electrochemical Impedance Spectroscopy technique under different operating conditions. We found that cathodes synthesized using smaller template pores exhibited better performance. Besides, SBCO cathodes displayed lower area-specific resistance than GBCO ones.

[*] Corresponding author. E-mail address: sacanell@tandar.cnea.gov.ar

## 1. Introduction

Extensive research has been devoted in the last few years to develop novel materials and microstructures for solid oxide fuel cells (SOFC) components to operate in the intermediate temperature (IT) range (500ºC - 700ºC) [1,2,3,4,5,6,7]. We have recently shown that the use of nanostructures is beneficial both for electrolytes [8,9] and electrodes [10,11,12]. In the case of the cathode, we developed nanostructured tubes and rods that show impressively low polarization resistance [13,14,15,16].

Typical SOFC cathodes are made mixed ionic and electronic conductors (MIEC) [17]. Among them, layered MIEC perovskites of formula $LnBaCo_2O_{6-\delta}$ (LnBCO, Ln = lanthanide) [18,19,20] have shown to display fast oxide-ion diffusion due to a reduction of the oxygen bonding strenght that leads to the appearance of channels for fast ion motion [21]. The layered ordering of cations, characterized by a sequence of atomic layers of Ln followed by a layer of Ba, is due to the large difference between ionic radii of $Ba^{2+}$ ($r_{Ba2+}$ ~ 1.61 Å), and $Ln^{3+}$ ions ($r_{Sm3+}$ ~ 1.24 Å, $r_{Gd3+}$ ~ 1.107 Å, for example).

Considering those results, it would be desirable to develop cathodes in order to combine both benefits, that related with the layered ordering and to the nanostructured character. However, so far and to the best of our knowledge, no studies have been performed on the study of nanostructured layered cathodes of LnBCO. The main reason may be due to the inherent difficulty to retain the nanostructure in layered perovskite materials, at the typically high temperatures needed for their synthesis procedure.

In this work we successfully obtained nanostructured LnBCO cathodes, with Ln = Gd and Sm, following a similar procedure developed to obtain nanostructured perovskites of different compositions [22,23,24]. By that method, we were able to retain both the nanostructured character and the layered structure.

The obtained materials were further evaluated as SOFC cathodes that display excellent electrochemical properties in the IT range. Thus, this work paves the way to develop other nanostructured layered perovskite materials in different research areas.

## 2. Experimental Procedure

We synthesized LnBCO nanostructures by using the pore wetting technique [22]. Commercial polycarbonate membranes with pore size of 200 and 800 nm were used as template and filled with a nitrate precursor solution. The membranes were further treated under microwave radiation for a 5 minutes and then calcined at 850 °C for 10 min). The resulting material was used as precursor for cathodes deposition.

XRD experiments were carried out using an Empyrean X-ray diffractometer from PANalytical B.V. (Laboratorio de Difracción de rayos X, Departamento Física de la Materia Condensada, Gerencia Investigación y Aplicaciones, GAlyANN, CAC-CNEA, Argentina) using Cu-K$\alpha$ radiation and PIXcel$^{3D}$ detector.

FE-SEM observations were performed using a FEI Helios NanoLab 650 (Centro de Investigación y Desarrollo en Mecánica, INTI, Argentina).

LnBCO nanostructures were smeared with a brush on both sides of $Ce_{0.8}Gd_{0.2}O_{1.9}$ electrolytes (GDC, Nextech Materials) to fabricate symmetrical [LnBCO/GDC/LnBCO] cells. The samples were dried at 50°C in air for about 20 min and sintered 1 h at 1050°C, with a heating and cooling rate of 10°C min$^{-1}$.

The Area-Specific Resistance (ASR) of the cathodes was determined from Electrochemical Impedance Spectroscopy (EIS) measurements performed with a Gamry 750 potentiost-galvanostat at zero bias. EIS measurements were performed in air, in pure oxygen and in a mixture of 95% of $N_2$ and 5% of $O_2$.

**3. Results and discussion**

XRD data for the LnBCO powders treated at 1050°C (the sintering temperature of the cathodes) are shown in Figure 1. We used the following nomenclature: Cathodes made with $SmBaCo_2O_6$ and templates of 200 nm and 800 nm were labelled as SBCO2 and SBCO8, respectively. Similarly, cathodes made with $GbBaCo_2O_{6-\delta}$ were labelled as GBCO2 and GBCO8. We show the data for samples obtained with templates of 200 nm of diameter, difractograms corresponding to cathodes obtained with templates of 800 nm present the same crystal structure (see Supplementary material). Single phase samples with orthorhombic crystal structure (*Pmmm* space group) were obtained for both compositions and for both template pore size studied in this work: 200 nm and 800 nm. Our XRD patterns are consistent with those presented in references [25] for Ln: Gd and [26] for Ln: Sm, indicating that both compounds present an ordered layered crystal structure. By analyzing the broadening of Bragg peaks, an average crystallite size of 40 nm was estimated using the Scherrer equation in all cases. This is an important result because, to the best of our knowledge, the electrochemical properties of nanostructured layered LnBCO have not been reported in the literature before.

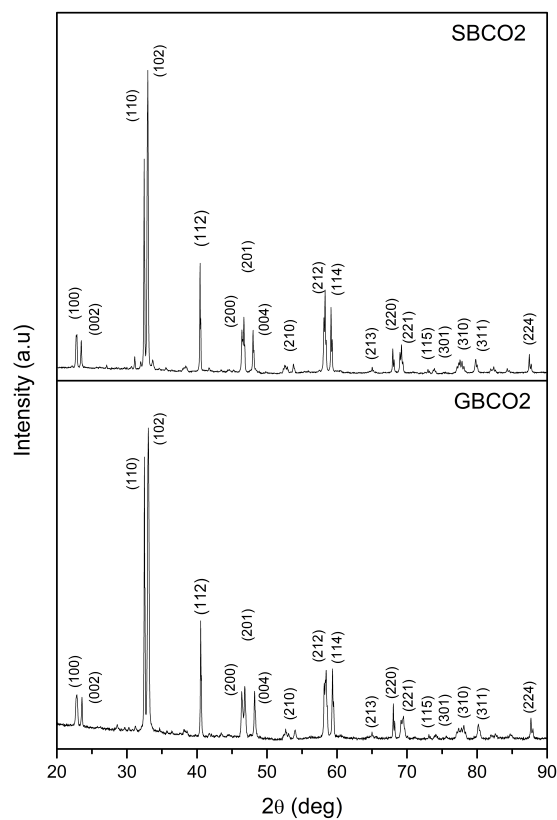

**Figure 1:** X-ray diffraction paterns for the samples SBCO2 and GBCO2 treated at 1050°C. Data corresponds to single phase samples with orthorhombic crystal structure of *Pmmm* space group.

SEM micrographs of the GBCO2 and GBCO8 cathodes are shown in figure 2. All cathodes are highly porous, but no signal of a tubular nor filamentary structure is observed, in contrast with previously observed results of other compositions [13,14]. We show micrographs corresponding to $GbBaCo_2O_6$ because both cathodes display similar characteristics, a complete set of pictures can be found on the Supplementary Material.

Cathodes made using templates of 200 nm pores are mainly formed by agglomerated particles of around 200 nm for both compositions although some larger particles are also observed. Cathodes made with templates of 800 nm pores are formed by particles of average diameter of 100 nm. The smaller diameter of the particles in the cathodes obtained from templates of 800 nm, is likely to be related to the small degree of agglomeration resulting in a larger confinement volume of the reactants.

Also, cathodes made with templates of 800 nm display two scales of porosity, they are porous both at the meso- and nanoscales. The former is due to the pore former of the ink vehicle and the latter being a result of the confinement performed by the templates.

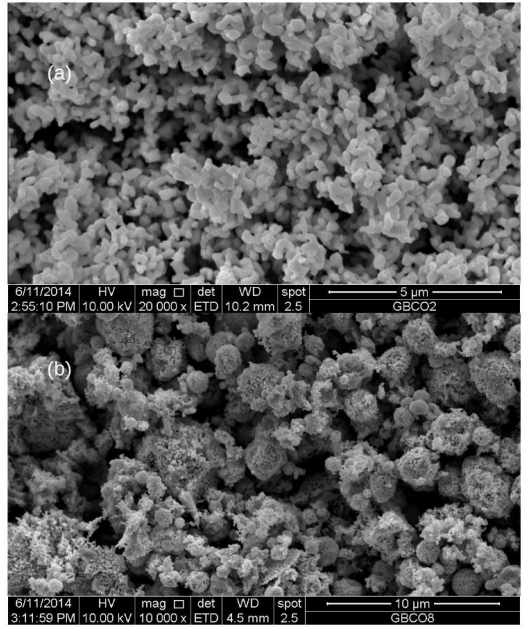

**Figure 2:** SEM micrographs of (a) the GBCO2 cathode and (b) the GBCO8 cathode. All cathodes are highly porous. Cathodes made using templates of 200 nm pores are mainly formed by agglomerated nanoparticles with no particular symmetry while cathodes made with templates of 800 nm pores are formed by nanoparticles agglomerated in "sphere-like" structures of micrometric scale.

The particles are distributed on larger structures which are reminiscent to porous spheres (for SBCO they are slightly distorted, see Supplementary Material). Those structures are the result of having used solutions of low concentration (to avoid precipitation of the reagents) in relatively large template pores.

The EIS spectra (in air, at 700°C and zero bias) of GBCO2 and GBCO8 is shown in figure 3 (a), while the corresponding spectra of SBCO2 and SBCO8 is shown in figure 3(b). All materials present promising cathodic properties at 700°C, reaching an ASR of 0.4 $\Omega$-cm$^2$ for the SBCO2 cathode. A significant influence with the diameter of the used precursors has been obtained for the cathode of GdBaCo$_2$O$_6$, in comparison with the cathode of SmBaCo$_2$O$_6$, being the latter the one displaying the lowest ASR. The lowest ASR of each composition is obtained for the precursors synthesized in templates of smaller pores.

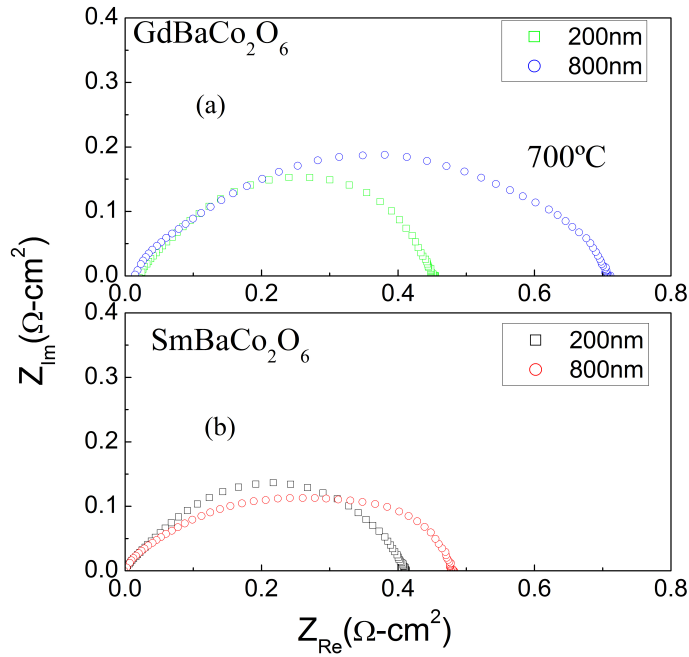

**Figure 3: (color online)** Electrochemical Impedance Spectroscopy data at 700°C for (a) GBCO2, GBCO8, (b) SBCO2 and SBCO8. The diameter of the used precursors has much significant influence in Gd-based cathodes than in Sm ones, but the latter displays the lowest ASR. The lowest ASR of each composition is obtained for the precursors synthesized in templates of smaller pores.

In figure 4 we show the Arrhenius plot of the ASR for all cases. We can see that almost all cathodes reach ASR values lower than 1 $\Omega\cdot cm^2$ above 650°C and of the order of 0.5 $\Omega\cdot cm^2$ at 700°C. In the inset we sketch the values of ASR as a function of each sample to analyze the influence of composition and nanostructure. At high temperatures cathodes SBCO2 and GBCO2 are clearly those with the best performance, showing that the microstructure dominates the cathodic behavior. On the other hand, on increasing temperature, composition seems to be much significant, as cathodes made with $SmBaCo_2O_6$ display lower ASR than cathodes of $GdBaCo_2O_6$ for fixed microstructure.

We found that the EIS spectra of our cathodes mainly consist two processes. One dominant process at intermediate and large frequencies, corresponding to a Warburg element as it is usual in cathodes with high oxide ion conductivity, in series with a parallel between a resistor ($R_1$) and a constant phase element($Q_1$) at low frequencies. No additional contributions were needed in order to fit the data. We confirmed that $R_W$ is related with oxide ion conduction and that $R_1$ is related with dissociative adsorption and gas phase diffusion, by their $p(O_2)$ dependence (see Supplementary Material).

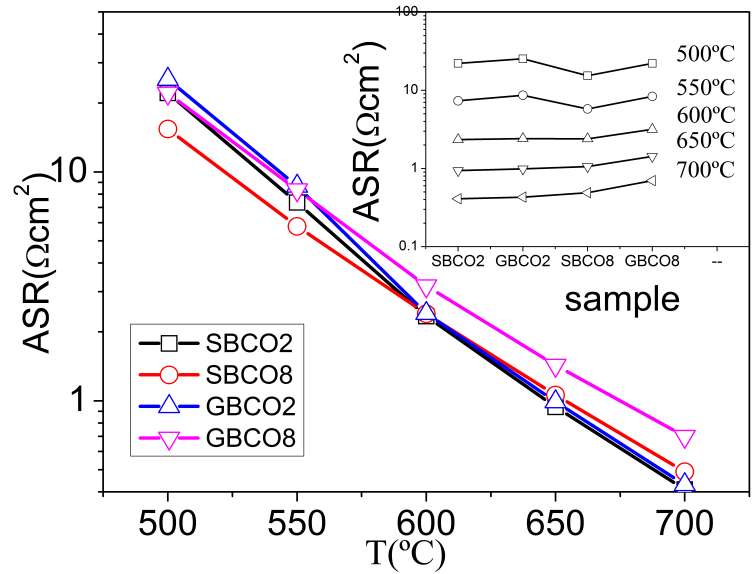

**Figure 4: (color online)** Area Specific Resistance as a function of temperature. Most cathodes reach lower ASR than 1 $\Omega.cm^2$ above 650ºC and of the order of 0.5 $\Omega$ -cm$_2$ at 700ºC. Inset: ASR as a function of the sample.

The resistive parts of the equivalent circuit obtained by fitting the data at 700ºC, are presented in table I (in the Supplementary Material we present the fitting of the EIS data at different oxygen partial pressures). In all cases, the resistive part of the Warburg element ($R_W$) represent the most significant part of the ASR. By the ratio $R_W/R_1$ we can compare the relative influence of each component. Comparing SBCO2 and SBCO8, with very similar values of $R_1$, its relative influence is major in SBCO2. This means that SBCO2 has a smaller resistance associated with diffusion. Such comparison is difficult between GBCO2 and GBCO8, because both resistive components differ for those cathodes, however, it is clear that the diffusion resistance ($R_W$) is smaller in GBCO2. A comparison between SBCO8 with GCBO8 evidences that ion conduction is improved in Sm-based cathodes.

|         | SBCO2 | SBCO8 | GBCO2 | GBCO8 |
|---------|-------|-------|-------|-------|
| $R_W$   | 0.28 $\Omega$-cm$^2$ | 0.39 $\Omega$-cm$^2$ | 0.41 $\Omega$-cm$^2$ | 0.6 $\Omega$-cm$^2$ |
| $R_1$   | 0.14 $\Omega$-cm$^2$ | 0.13 $\Omega$-cm$^2$ | 0.04 $\Omega$-cm$^2$ | 0.14 $\Omega$-cm$^2$ |
| $R_W/R_1$ | 0.5 | 0.33 | ~0.1 | 0.23 |

**Table I:** Resistive parts of the fitting components. $R_W$ (corresponding to intermediate to high frequency the Warburg element) and $R_1$ (corresponding to the low frequency $R_1Q_1$ parallel).

## 4. Conclusions

In summary, we obtained nanostructured layered perovskites of $SmBaCo_2O_6$ and $GbBaCo_2O_{6-\delta}$ by a very simple procedure and attached to an electrolyte retaining its original nanostructural character. The obtained cathode is highly porous due to the precursors powder. All cathodes display very low ASR in the IT range, moreover taking into account that they were simply smeared with a brush for deposition. Our results show that smaller scale nanostructures (accomplished by using templates of smaller diameter) display an enhanced performance. The main contribution to ASR is bulk diffusion of oxide ions, with a minor low frequency contribution ascribed to the redox process in the electrode surface and gas phase diffusion.

Besides the promising results in the area of IT-SOFCs, the successful synthesis of nanostructured layered perovskites, paves the way for their use in other applications such as magnetoresistive materials or materials for solar cells.

## Acknowledgements

The present work was partially supported by Agencia Nacional de Promoción Científica y Tecnológica (Argentina, PICT 2011 No. 1948) and CONICET (Argentina, PIP 000362-2011).